\documentclass[aps,prl,twocolumn,showpacs]{revtex4}

\usepackage{psfig}

\begin{document}

\title{Electron-phonon effects on the Raman spectrum in MgB$_2$}

\author{E. Cappelluti}

\affiliation{``Istituto dei Sistemi Complessi'', CNR-INFM,
v. dei Taurini 19, 00185 Roma, Italy,}

\affiliation{and Dipart. di Fisica, Universit\`{a} di Roma ``La Sapienza",
P.le A. Moro, 2, 00185 Roma, Italy}

\date{\today}

\begin{abstract}
The anomalous features of the Raman spectroscopy measurement in MgB$_2$
represent a still unresolved puzzle.
In particular highly debated are the origin of the huge $E_{2g}$ phonon
linewidth, the nature of the low energy ($\omega < \omega_{E_{2g}}$)
background and the evolution of the Raman spectra with Al doping.
In this paper we compute the self-energy of the $E_{2g}$ phonon mode
in a fully self-consistent way
taking into account electron-phonon effects on the electronic properties.
We show that all the anomalous features can be naturally understood
in a framework where the whole electron-phonon spectrum
$\alpha^2F(\omega)$ gives rise to significant damping processes
for the electronic excitations and consequently
for the $E_{2g}$ phonon itself. The two-peak structure as function
of the Al doping is ascribed to finite bandwidth effects arising
as the Fermi level approaches the $\sigma$ band edge.
\end{abstract}
\maketitle

There is nowadays a general consensus that the
electron-phonon (el-ph) interaction is responsible for the superconductivity
in MgB$_2$ \cite{an,kortus,yildirim,liu,kong,choi}.
Interesting peculiarities make however this compound
quite unique and different from the conventional low-$T_c$ superconductors.
Most notable are the two-bands/two-gaps
phenomenology \cite{liu,choi,Bouquet,Szabo,Giubileo,Tsuda,Gonnelli}, the fact
that a large part of the electron-phonon coupling is concentrated
in just one phonon mode $E_{2g}$ \cite{an,yildirim,kong},
the possibility of spanning
with Al doping a wide range of hole doping of the $\sigma$
bands \cite{slusky}.
Although several features are now well understood
within the context of the two-gaps scenario, the experimental
overview of the Raman spectroscopy of this compound still represents
a highly puzzling
anomaly \cite{bohnen,hlinka,goncharov,postorino,renker,quilty,martinho}.
The main open questions on this point
regard in particular the origin of the anomalously large linewidth
of the $E_{2g}$ phonon mode, the nature of the background Raman
signal at low frequencies, and the evolution of the Raman spectra
upon Al doping, which shows a transfer of spectral weight between
different peaks more than a continuous hardening of the $E_{2g}$ phonon mode.

In this Letter we show that all these unconventional features can be
explained within the electron-phonon scenario by the coexistence
of a strongly coupled ($E_{2g}$) phonon mode at ${\bf q}=0$ and
of a wide background electron-phonon scattering with other phonon branches
and different momenta. This gives rise to a Fano-like phenomenology
where the shape and linewidth of the $E_{2g}$ mode is mainly related
to the electron-phonon decay processes triggered by the other modes
and by impurity scattering. Similar effects give rise to a
low energy peak, equivalent to the Drude-like feature in the optical
conductivity, which is responsible for the low energy signal.
We show also that the transfer of spectral weight between different
Raman structures as function of the Al amount can be a direct consequence
of the vanishing of the Fermi energy.

The origin of the broad phonon linewidth in Raman spectroscopy
has been widely discussed in
literature \cite{bohnen,renker,shukla,lazzeri,calandra}.
Common feeling on this subject is that the linewidth of a phonon mode
$\gamma_{\bf q}$ reflects the strength of the electron-phonon
coupling $\lambda_{\bf q}$ of this specific mode,
in the spirit of the Allen's formula
$\gamma_{\bf q}=2\pi N(0) 
\lambda_{\bf q}\omega_{\bf q}^2$ \cite{allen},
where $N(0)$ is the electron density of states (DOS) coupled
with the phonon mode and $\omega_{\bf q}$ is the phonon frequency.
First-principle calculations however indicate that the electron-phonon
coupling alone of the $E_{2g}$ phonon mode is not sufficient to explain
the large experimental Raman linewidth \cite{calandra}.
In addition, the direct employment of the Allen's formula is strongly
questioned in the ${\bf q}=0$ case, relevant for the
Raman spectroscopy, where a more careful analysis shows that
the damping of a ${\bf q}=0$ phonon mode scattering with a
{\em non-interacting} electronic system is strictly zero \cite{calandra}.

In contrast to the above scenario,
a Quantum Field Theory analysis specifically addressed to the 
${\bf q}=0$ case has been developed
in Refs. \cite{zeyher,marsiglioph}. Although these studies were
mainly aimed to investigate phonon anomalies
in the superconducting state, this approach can be as well employed in
the normal state. Along this line, for instance,
Marsiglio {\em et al.} remarked \cite{marsiglioph,zawadowski}
that the imaginary part of a ${\bf q}=0$ phonon self-energy
in the weak-coupling limit is simply
\begin{equation}
\Pi''({\bf q}=0,\omega) \propto 
\omega_{\bf q} \lambda_{{\bf q}=0}
\frac{4 \omega \Gamma_{\rm imp}}{\omega^2+4\Gamma_{\rm imp}^2},
\label{imp}
\end{equation}
where the impurity scattering rate $\Gamma_{\rm imp}$ was assumed
to be the only damping source of the electronic propagator:
$G({\bf k},\omega)=1/[\omega-\epsilon_{\bf k}+i\Gamma_{\rm imp}]$.

Eq. (\ref{imp}) explicitly shows that the damping processes of a ${\bf q}=0$
phonon are triggered by corresponding damping processes of the electronic
charge response which screens the phonon, whereas the el-ph coupling
$\lambda_{{\bf q}=0}$ of this particular mode rules the magnitude
of these effects. This results holds true also
when the main source
of the the electronic excitation damping
is the electron-phonon interaction itself.
In this case it is also important to note that in principle
all the phonon modes, and not only the specific $\lambda_{{\bf q}=0}$
mode, contribute to the electronic damping.
Indeed, as we are going to discuss,
the imaginary part $\Pi''(\omega)$
of the phonon self-energy is roughly related to the
integral up to the energy $\omega$
of the imaginary part of the {\em electronic} self-energy,
$
\gamma({\bf q}=0,\omega) 
\propto 
\lambda_{{\bf q}=0}
\int_0^{\omega} d\omega' \Gamma(\omega')$.
For optical phonons close to the top
of the phonon spectrum  we have $\int_0^{\omega} d\omega' \Gamma(\omega')
\propto \lambda \omega$, $\lambda$ being the {\em total}
electron-phonon coupling constant, so that
in intermediate-strongly coupled systems
the phonon linewidth can be of the same order of
the phonon frequency itself.
In conventional low-$T_c$ superconductors however the total
electron-phonon coupling $\lambda \sim 1$ is spread over several
phonon branches and ${\bf q}$ modes,
$\lambda = \sum_{{\bf q},\nu} \lambda_{{\bf q},\nu}$,
so that the contribution from
a single mode is quite small.
Typical values of $\lambda_{{\bf q}=0,\nu}$ for the
$A_g$ phonon modes in cuprates are
for instance $\lambda_{{\bf q}=0,\nu} \le 0.034$ \cite{rodriguez},
resulting in experimental Raman linewidths of few meV \cite{friedl}.
The opposite extreme case of only one strong coupled
${\bf q}=0$ phonon mode $\omega_0$
carrying the whole electron-phonon coupling would also predict
extremely small phonon linewidth since,
although $\lambda_{{\bf q}=0} \sim 1$,
the amount of electron-phonon coupling smaller than $\omega_0$
is practically negligible and 
$\int_0^{\omega_0} d\omega \Gamma(\omega) \approx 0$.

\begin{figure}
\centerline{\psfig{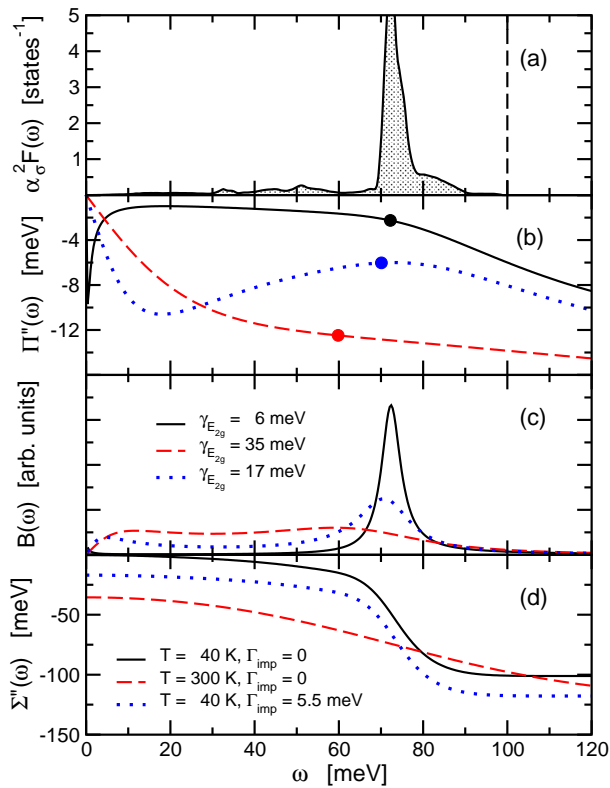}}
\caption{(color online) 
(a) Electron-phonon
spectral function $\alpha^2_\sigma F(\omega)$ (from Ref. \cite{golubov}).
The dashed
line represents the unrenormalized phonon frequency
$\Omega_{E_{2g}}=100$ meV.
Panels b,c,d show respectively the
imaginary part of the phonon self-energy,
the phonon spectral function and
the imaginary part of the electronic self-energy
for three representative cases.
}
\label{f-tot}
\end{figure}

Along this scenario MgB$_2$ presents very peculiar characteristics
since it presents at the same time two (degenerate)
strong coupled ${\bf q}=0$ $E_{2g}$
phonon modes {\em and} a relevant fraction of the total electron-phonon
coupling strength spread over other different
modes \cite{an,yildirim,kong,choi}.
We employ a fully self-consistent many-body approach to investigate
the normal state $E_{2g}$ Raman spectrum of MgB$_2$. In more explicit way we
first solve iteratively the Marsiglio-Schossmann-Carbotte
equations \cite{msc}
with the Eliashberg spectral function $\alpha^2_\sigma F(\omega)$
obtained by first-principle calculations \cite{golubov}
(Fig. \ref{f-tot}a) to evaluate the real-axis
electronic self-energy $\Sigma(\omega)$ of the $\sigma$ bands.
The Eliashberg function $\alpha^2_\sigma F(\omega)
=\alpha^2_{\sigma\sigma} F(\omega)+\alpha^2_{\sigma\pi} F(\omega)$
describes the total electron-phonon scattering of the $\sigma$ electrons
with both the $\sigma$ ad $\pi$ bands.
Note that the $\alpha^2F(\omega)$
extracted from  first-principle techniques represents the el-ph
spectral function where phonons are {\em already} renormalized.
In order to better compare with the experiments, we include also explicitly
possible effects of impurity disorder.
We schematize the $\sigma$-bands as two degenerate bands
with constant DOS, namely:
$\sum_{\bf k} G({\bf k},\omega) = N_\sigma(0) \int d\epsilon
G(\epsilon,\omega)$, where $N_\sigma(0)$ represents the $\sigma$-band
electron density of states per spin and per band, $N_\sigma(0)\simeq 0.075$
states/(eV $\cdot$ cell) \cite{an,kortus}.
The so-obtained electronic Green's function is then employed as input
to calculate the full frequency dependence of the
${\bf q}=0$ phonon self-energy $\Pi(\omega)$ \cite{marsiglioph},
whose imaginary part reads:
\begin{eqnarray}
\Pi''(\omega)
&=&\frac{\pi N_s N_b}{N_c}\sum_{\bf k}
|g_{{\bf k},E_{2g}}|^2 \int d\omega'
A({\bf k},\omega'+\omega)
\nonumber\\
&&\times
A({\bf k},\omega')
\left[
f(\omega'+\omega)-f(\omega')
\right],
\label{s_ph}
\end{eqnarray}
where $N_c$ is the number of sampling point in the Brillouin zone,
$A({\bf k},\omega)=(1/\pi)
\mbox{Im}[1/(\omega-\epsilon_{\bf k}-\Sigma(\omega))]$
is the electron spectral function and
$N_s=2$ and $N_b=2$ represent respectively the spin degeneracy
and the $\sigma$-band degeneracy. Note that Eq. (\ref{s_ph})
does not account the double degeneracy of the two $E_{2g}$ modes
which on the other hand contribute to the superconducting pairing.
A similar expression is obtained for the real part of the phonon
self-energy, also attainable from the Kramers-Kr\"onig relations.
The electron-phonon matrix elements
$g_{{\bf k},E_{2g}}$ can be estimated from the $E_{2g}$
deformation potential 
$D_{E_{2g}}\simeq 12$ eV/\AA \cite{an,yildirim,boeri,profeta},
which is in good approximation ${\bf k}$-independent close
to the Fermi level, and from evaluating the zero point motion lattice
displacement $g_{E_{2g}}= D_{E_{2g}} \sqrt{\langle u^2 \rangle}= 0.39$ eV,
taking into account anharmonic effects \cite{boeri2}.

We finally evaluate
the phonon spectral function $B(\omega)$
from the imaginary part of the phonon propagator
\begin{equation}
B(\omega)=-\frac{1}{\pi}\mbox{Im}
\left[
\frac{2\Omega_{E_{2g}}}{\Omega_{E_{2g}}^2-\omega^2
+2\Omega_{E_{2g}}\Pi(\omega)}
\right],
\end{equation}
where $\Omega_{E_{2g}}$ is the {\em unrenormalized} phonon frequency,
while the renormalized phonon
frequency $\omega_{E_{2g}}$ and the phonon linewidth $\gamma_{E_{2g}}$
are simply related to the  real and imaginary parts of $\Pi(\omega)$,
\begin{eqnarray}
\omega_{E_{2g}}^2
&=&
\Omega_{E_{2g}}^2+2\Omega_{E_{2g}}\Pi'(\omega_{E_{2g}}),
\label{wren}
\\
\gamma_{E_{2g}}&=& -2(\Omega_{E_{2g}}/\omega_{E_{2g}})
\Pi''(\omega_{E_{2g}}).
\label{gamma}
\end{eqnarray}
We assume the unrenormalized
$E_{2g}$ phonon frequency $\Omega_{E_{2g}}=100$ meV,
close to the top of the phonon spectrum.

In Fig. \ref{f-tot}b,c we plot the imaginary part of the phonon
self-energy, and the fully renormalized phonon spectrum.
The filled symbols mark the values of
imaginary parts of the phonon self-energy 
evaluated at the renormalized phonon frequency $\omega_{E_{2g}}$
obtained from the self-consistent solution of Eq. (\ref{wren}).
At low temperature and in the absence of impurity scattering,
the imaginary part of the phonon self-energy shows a monotonic behavior
which reflects the corresponding increasing of the
electronic damping processes (Fig. \ref{f-tot}d). Already in this case
we predict a relevant value of $\Pi''(\omega_{E_{2g}})$
and of the phonon linewidth  $\gamma_{E_{2g}} \simeq 6$ meV, which
is essentially only due to the spectral weight of
$\alpha^2_\sigma F(\omega)$ for $\omega\le \omega_{E_{2g}}$.
Things are even more drastic when finite temperature effects
or impurity scattering are taken into account.
As is well known the imaginary part of the electronic self-energy
starts from a finite value at $\omega=0$ which is reflected in a sudden
increase of $|\Pi''(\omega)|$ and in a significant broadening
of the phonon spectrum.
Note also the appearance of an incoherent background for 
$\omega\le \omega_{E_{2g}}$ and of a broad shoulder at very low temperature,
in striking agreement with the Raman experimental data.
Similar effects were pointed out in Refs. \cite{zawadowski,dahm}
in the context of the electronic Raman scattering.
We would like to stress that the sharp increase of
$|\Pi''(\omega)|$ and the onset of the low temperature shoulder
in the phonon spectrum are intrinsic features of the
charge response function. As a matter of fact,
as noted in Ref. \cite{marsiglioph},
at a first approximation
the ratio $-\Pi''(\omega)/\omega$ is qualitatively similar
to the real part of the optical conductivity $\sigma'(\omega)$.
The low energy sudden enhancement of $|\Pi''(\omega)|$ and the corresponding
shoulder in the phonon spectrum, which arise from the finite
value of $\Sigma''(\omega=0)$, have thus a strict connection
with the appearance of a Drude-like peak in $\sigma'(\omega)$.
Along this line, the experimental observation of a significant Drude-like
scattering rate $1/\tau \simeq 9-37$ meV \cite{tu,kaindl}
and of the Raman shoulder
at low temperatures $T \sim 40-45$ K \cite{quilty} points out the actual
presence
of  a small amount of impurity scattering.
From the comparison between our results and experimental data
we estimate an impurity scattering rate $\Gamma_{\rm imp}\simeq 5.5$ meV.

\begin{figure}
\centerline{\psfig{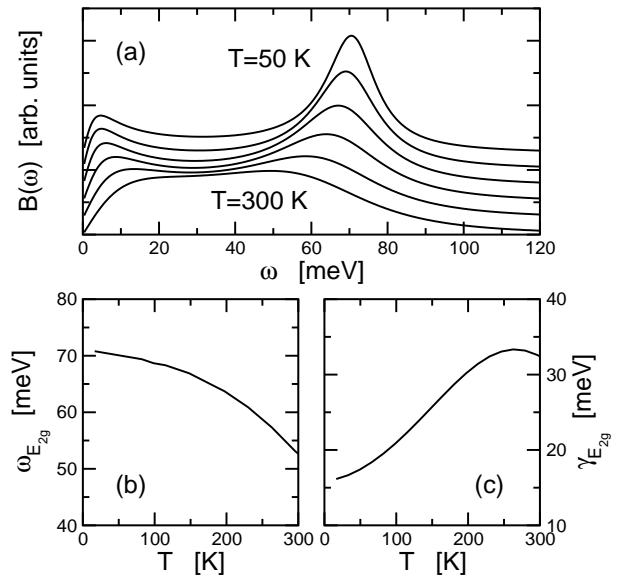}}
\caption{(a) Phonon spectral function for $T=50, 100, 150, 200, 250, 300$ K;
(b) temperature dependence of the renormalized phonon frequency 
$\omega_{E_{2g}}$; (c) temperature dependence of the phonon linewidth
$\gamma_{E_{2g}}$. We considered here a small amount of impurity concentration
corresponding to $\Gamma_{\rm imp}= 5.5$ meV.}
\label{f-spectra}
\end{figure}

In Fig. \ref{f-spectra}a we show the evolution of the phonon spectral
function $B(\omega)$ at function of the temperature. 
Increasing temperature leads thus not only to a smearing of the
main phonon peak by increasing $\gamma_{E_{2g}}$ but also to a smearing
of the low energy structure, in fair agreement with the experiments.
In Fig. \ref{f-spectra}b,c we show also the temperature dependence
of the renormalized phonon frequency $\omega_{E_{2g}}$
and of the phonon linewidth $\gamma_{E_{2g}}$. The agreement with
the experimental data is once more remarkable for 
$\gamma_{E_{2g}}$ \cite{martinho}.
On the other hand our analysis underestimates the
phonon frequency $\omega_{E_{2g}} \sim 55-70$ meV
(to be compared with the $\omega_{E_{2g}} \sim 78$ meV from Raman
spectroscopy) and predicts a phonon softening
as a function of $T$ which has not been observed. We remind however that
anharmonic effects are not taken into account here, and that they would
lead to a temperature dependent hardening of the
$E_{2g}$ phonon frequency. Since the anharmonic effects were shown
to be related to the eletron-phonon interaction itself \cite{boeri},
a correct evaluation of them would require once more
the inclusion in a consistent way
of the damping electronic processes.

As a last point of our analysis we address the evolution of the Raman
spectra as function of the Al doping. Ab-initio calculations predict
that the electron doping induced by the Al content would fill
the two $\sigma$ bands, leading to a reduction of the electron-phonon
coupling. This is expected to result in a steady hardening of
the $E_{2g}$ phonon frequency \cite{profeta}. 
Actual Raman measurements are however
in substantially disagreement with this picture, showing that
the phonon hardening occurs via a transfer of spectral weight from
the $E_{2g}$ peak to another, at the moment unknown,
new high energy structure \cite{renker,postorino}. 
In the following we provide a qualitative
scenario where this behavior
is naturally explained in terms of finite bandwidth effects which arise when
the Fermi level approaches the $\sigma$ band edge.

\begin{figure}
\centerline{\psfig{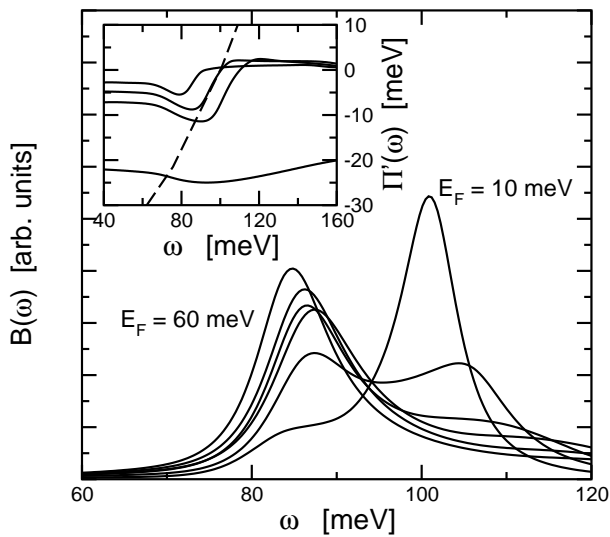}}
\caption{Evolution of the phonon spectral function by decreasing
the $\sigma$ band Fermi energy for $T=40$ K and $\Gamma_{\rm imp}=0$.
Inset: real part of the phonon self-energy (solid lines)
for different $E_{\rm F}$, from top to bottom:
$E_{\rm F}=10, 20, 30, \infty$ meV. The crossing with the dashed line
determines the phonon peaks.}
\label{f-dop}
\end{figure}

A natural modelization to investigate this scenario would be to consider
a semi-infinite band system  $\int_{-\infty}^\infty d\epsilon \rightarrow 
\int_{-\infty}^{E_{\rm F}}$, where the Fermi energy $E_{\rm F}$
plays the role of an energy cut-off for the electronic excitations.
However the correct treatment of the charge conservation and
of the asymptotic behavior at $\omega \rightarrow -\infty$ in the real axis
phonon self-energy\cite{marsiglioph} present different numerical difficulties
which make this approach unaffordable. 
For a qualitative insight we consider thus
in the following a simpler model with a symmetric band
$\int_{-E_{\rm F}}^{E_{\rm F}}$ where these numerical problems are overcome.
Since the main role is played by the presence of a finite cut-off $E_{\rm F}$,
this model is expected to shed a qualitative light also on more realistic
cases.
In Fig. \ref{f-dop}
we show the evolution of the phonon spectral function
for $E_{\rm F}$ approaching zero for $T=40$ K and $\Gamma_{\rm imp}=0$.
For $E_{\rm F}$ higher than 60 meV (not shown here)
the phonon spectrum presents only a one-peak structure
corresponding to the renormalized phonon frequency.
As soon as $2E_{\rm F}$ decreases,
however, this structure rapidly looses spectral weight in favor
of the unrenormalized phonon frequency $\Omega_{E_{2g}}$
accompanied by a narrowing of the phonon linewidth.
This behavior can be understood by looking at the real part
of the phonon self-energy: for $E_{\rm F}\rightarrow \infty$
it has a negative minimum at roughly the maximum energy
of $\alpha_{\sigma}^2F(\omega)$, and then it vanishes very slowly
for $\omega \rightarrow \infty$.
Peaks in the phonon spectrum
are qualitatively determined by Eq. (\ref{wren}), corresponding to the
crossing of $\Pi'(\omega)$ with the dashed line 
[$(\omega^2-\Omega_{E_{2g}}^2)/2\Omega_{E_{2g}}$]
in inset of Fig. \ref{f-dop}.
The finite band edge $E_{\rm F}$ is reflected in a sharp drop
of $\Pi'(\omega)$ due to the cut-off in the particle-hole excitation,
with a width roughly given by $E_{\rm F}$. 
Decreasing $E_{\rm F}$ the determination of $\omega_{E_{2g}}$
according Eq. (\ref{wren}) has an abrupt jump from a renormalized
($\Pi'(\omega)\neq 0$) to the unrenormalized ($\Pi'(\omega)\simeq 0$)
value. The crossover between these two regimes occurs roughly at
$E_{\rm F} \simeq 20$ meV where Eq. (\ref{wren}) has three solutions,
corresponding to the two peaks and the middle minimum in the
phonon spectrum.

We have been unable to show evidence of this behavior
even at room temperature and in the presence of
impurity scattering $\Gamma_{\rm imp}=6$ meV, where
the large broadening of the phonon spectrum does not permit
to resolve a two peak structure. One has to take into in account however
that approaching the band edge also the electron DOS and the
impurity scattering rate $\Gamma_{\rm imp} \propto N(0)$
vanishes. The resemblance of this
framework with the experimental results \cite{renker,postorino}
is anyway striking suggesting that
finite bandwidth effects are actually the natural explanation
of the evolution of the Raman spectrum with Al doping.

In conclusion in this paper we have investigated
the phonon Raman spectroscopy data by computing
the phonon self-energy of the $E_{2g}$ mode.
We show that the anomalous features of the Raman measurements,
namely the huge phonon linewidth, the low energy background,
the two-peak structure as function of the Al doping, can be naturally
explained by the interplay of the $E_{2g}$ phonon mode
with the whole electron-phonon spectrum which gives rise
to damping processes in the electronic excitation and in the
$E_{2g}$ mode itself.

We thank L. Pietronero, G.B. Bachelet,
P. Postorino, M. Lavagnini and D. Di Castro,
for interesting discussions.
We also acknowledge financial support
from the MIUR projects FIRB RBAU017S8R and COFIN 2003.

\end{document}